\documentstyle[11pt,paspconf]{article}
\begin{document}

\title{TRIPLETS OF GALAXIES: Some Dynamical Aspects}

\author{H\'ector Aceves}
\affil{IAA-CSIC. Apdo. Postal 3004. Granada 18080, Spain. $\;${\sf aceves@iaa.es}}



\section{Introduction}

In celestial mechanics the 3-body problem has a long and rich history, while the  problem of 3-galaxy systems is rather scarce (see review by Valtonen \& Mikkola 1991). Most of the up-to-date works have addressed the 3-galaxy problem using a point-like approach, although some explicit-physics simulations have been performed to simulate dynamical friction effects and merging processes (Zheng, et al. 1993).  These studies have provided important knowledge on the general behaviour of triplets of galaxies that are observed in the sky (Karachentsev 1999), and some accordance with observations has been obtained.

However galaxies are not point-like particles, but rather consist of a large number of stars that in turn can be approximated as point-like particles. Qualitative and quantitative differences result in the dynamics when the 3-galaxy problem is addressed self-consistently; {\it i.e.}~when galaxies are able to re\-distribute energy and angular momentum among their stars. Since not using self-consistent galaxies casts necessarily some doubts on earlier results,  we address here the 3-galaxy problem self-consistently and compare some results with observations.  A full report on this work is in preparation.

\section{Numerical Experiments}

Galaxies were modeled after a Plummer sphere with $N=3000$ particles each.  No explicit difference was made as to particles being luminous or dark. The units used here are such that $G=M=R_0=1$, where $M$ is the mass of each galaxy and $R_0$ its scale-length. To transform $N$-body results (`n') to astronomical ones (`a') we need to choose a real galaxy. We use here a galaxy similar to ours with $M\approx 5.5 \times 10^{11}$ M$_\odot$ and $R_{\rm halo}\approx 135$ kpc (Kuijken \& Dubinski 1995, Model B). The following transformations follow:
\begin{equation}\label{eq:units}
\frac{r_{\rm a}}{r_{\rm n}}  \approx 13.5 \, {\rm kpc} , \; \;
\frac{m_{\rm a}}{m_{\rm n}} \approx  5.5 \times 10^{11} \, {\rm M}_\odot , \;\;
\frac{t_{\rm a}}{t_{\rm n}} \approx 32 \,{\rm Myr} , \;\;
\frac{v_{\rm a}}{v_{\rm n}} \approx 420 \, {\rm km}/{\rm s} \,.
\end{equation}

The initial positions of the centre-of-mass of galaxies were sampled from a homogeneous spherical distribution of radius $R_{\rm max}$. The evolution of the triplet depends, obviously, on the size of this sphere.
This radius is taken here as an approximate turn-around radius of a density perturbation with the mass of a triplet; galaxies are assumed to be already formed. 
	We consider an initial virial ratio of $2T/|W|=1/4$ for this perturbation with velocity dispersion $ \sigma = V_0/2\sqrt{3},$ where
$V_0 = \sqrt{3 G M_{\rm t}/5 R_{\rm max}}\,$
and $M_{\rm t}=3M$ (e.g. Barnes 1985). 
	Numerical simulations of galaxy formation  tend to indicate that the dark matter background at turn-around had $\sigma \sim 20$ km/s (Lake \& Carlberg 1988). We take this as a fiducial value for $\sigma$ in triplets at maximum expansion. Hence for three Galaxy-like spirals we obtain $R_{\rm max} \sim 1$ Mpc.
	No common dark matter halo has been used in the present simulations.  Note that if we increase the mass of the triplet by introducing a common dark matter halo, and retain cold initial conditions, this will increase $R_{\rm max}$ proportionally.  The collapse time is taken here as $\tau_{\rm coll} = \pi \sqrt{R_{\rm max}^3/ 2 G M_{\rm t}} \approx 794 \approx 20$ Gyr. These IC's are more appropiate for triplets turning around at this epoch.

	We also considered virialized initial conditions (IC's), obtained by assuming galaxies to be point-masses. We made 30 simulations and estimated the 1-D velocity dispersion $\sigma$, mean harmonic radius $R_{\rm H}$, crossing time $t_{\rm c}$, and {\sl virial} mass $M_{\rm v}$ (Nolthenius \& White 1987),  along three orthogonal projections; {\it i.e.} 90 `triplets' are simulated for each of the IC's considered.  Simulations lasted $\approx 20$ Gyr from turn-around, and energy was conserved for all the runs to $\la 1.5$\%.

\section{Results} 

\begin{figure}[!t]
\plotfiddle{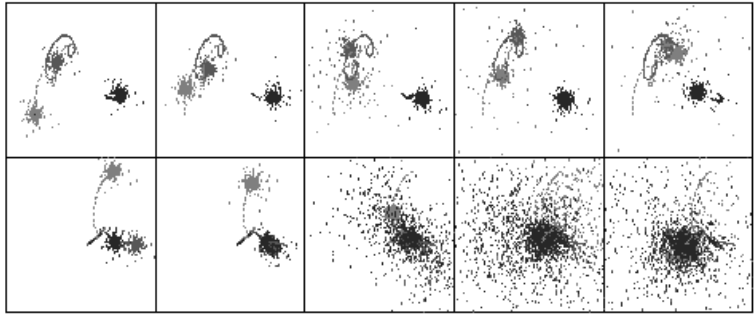}{4cm}{0}{60}{60}{-145}{-10}   
\caption{Outcomes of a triplet simulation, from initial conditions to $\approx 20$ Gyr in steps of $\approx 5$ Gyr. ({\it top}) A binary and a single galaxy results, ({\it bottom}) a triple merger is produced. Lines indicate the path of the centre of galaxies. Frames are $100\times 100$ units.}
\label{fig-1}
\end{figure}

In Figure~\ref{fig-1} we show two typical outcomes of the simulations performed. Qualitatively situations where binaries are formed first, some of them leading to a rather quick triple merger ({\it bottom}) and others taking a much longer time to even form a binary merger ({\it top}), are predominantly found.  This resembles the same instability found in the 3-body problem to form initially a binary.  However, when self-gravity is considered effects such as a sling-shot are not easy to reproduce due to the galaxies' capability to absorb orbital energy.

In Table~\ref{tab-1} we present quantitative results for both types of IC's, taking out mergers. Numbers are given in $N$-body units. The first row, for each time, are average values while in the second median values are given. Times correspond respectively to $\approx$ 0.5, 5, 10, 13, 15, and 20 Gyr.
As expected, initially virialized triplets evolve much slower than cold ones.
Collapsing triplets yield a significant number of mergers in the time interval $(10-13)$ Gyr; {\it i.e.}~$\sim \tau_{\rm coll}/2$.
	The average or median mass values {\it never} overestimate of the true mass of the system; in collapsing triplets.  The underestimate in the median is a factor of $\la 3$ than the true mass during the wide time interval of $\approx (5-15)$ Gyr for collapsing triplets; the agreement is much better for initially viralized systems. Some triplets provided an overestimate in mass along a particular line-of-sight.

\begin{table}
\caption{Results for Triplets in Isolation}
\label{tab-1}
\begin{center}\scriptsize
\begin{tabular}{r|rrrr|rrrr}
 \multicolumn{1}{c}{} & \multicolumn{4}{c}{Cold-Collapse} & \multicolumn{4}{c}{Virial} \\
t & $\sigma$ & $R_{\,\rm H}$ & $t_{\rm c}/\tau_{\rm coll}$ & $M_{\rm V}$ &  $\sigma$ & $R_{\,\rm H}$ & $t_{\rm c}/\tau_{\rm coll}$ & $M_{\rm V}$ \\
\hline
17.24  & 0.036 & 82.86 & 3.70 & 0.58  & 0.071 & 82.92 & 1.85 & 2.28 \\
       & 0.033 & 86.35 & 3.75 & 0.59  & 0.067 & 86.03 & 1.81 & 2.28 \\
159.42 & 0.063 & 73.21 & 2.91 & 1.64  & 0.102 & 73.93 & 1.41 & 4.08 \\
       & 0.050 & 71.06 & 1.71 & 1.04  & 0.087 & 71.95 & 1.28 & 3.46 \\
314.54 & 0.098 & 59.66 & 1.90 & 2.26  & 0.090 & 75.71 & 1.67 & 3.45 \\
       & 0.090 & 50.12 & 0.86 & 1.45  & 0.077 & 75.36 & 1.39 & 3.13 \\
409.33 & 0.091 & 58.26 & 1.88 & 2.18  & 0.086 & 74.38 & 1.77 & 3.39 \\
       & 0.065 & 52.62 & 1.71 & 1.65  & 0.082 & 68.39 & 1.41 & 2.82 \\
469.65 & 0.105 & 49.76 & 1.31 & 2.46  & 0.090 & 71.90 & 1.92 & 3.68 \\
       & 0.081 & 39.22 & 0.58 & 2.07  & 0.074 & 67.44 & 1.27 & 2.41 \\
642.00 & 0.066 & 43.27 & 1.62 & 1.04  & 0.058 & 67.71 & 2.45 & 1.32 \\
       & 0.065 & 30.88 & 0.70 & 0.84  & 0.054 & 63.09 & 1.50 & 1.17 \\
\hline
\end{tabular}
\end{center}
\end{table}

Results for the median $R_{\rm H}$ in collapsing triplets do not agree well with those of observations of present day compact triplets ($\approx 65$ kpc). This also happens for virialized systems which have a $R_{\rm H} \ga 400$ kpc. On other hand, the velocity dispersions are always $\sigma \la 50$ km/s for both types of IC's, a value which is about half of the observed median ($\approx 100$ km/s).  Nonetheless, about 10\% of the simulated triplets have $\sigma \ga 100$ km/s at $t\sim 10$ Gyr. Using a larger mass and halo size galaxy does not help much in increasing the astronomical median-$\sigma$ since velocities scale as $v \propto \sqrt{M/R}$.

We recall that when comparing to observational data the assumption that Kara\-chentsev's catalog of compact triplets forms an homogeneous and well defined sample is implicit, but this is not so since e.g.~it includes galaxies of different luminosities ($\sim$ mass) and morphological type. This needs to be considered in future studies to make  more consistent comparisons with observations.

\begin{figure}
\plotfiddle{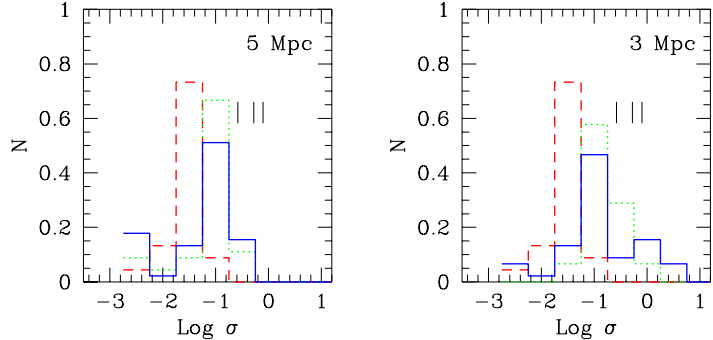}{4cm}{0}{50}{50}{-150}{-85}
\caption{Velocity dispersion distribution for collapsing triplets under the action of tidal perturbation from a point-mass $M\approx 10^{14}$M$_\odot$. ({\it left}) Triplet at 5 Mpc away from the perturber, and ({\it right}) at 3 Mpc. Dashed-lines are at $t=0$, dotted at $\approx 10$ Gyr, and solid-line at $\approx 15$ Gyr. Bars correspond to $\sigma \approx 100, 200, 300$ km/s, respectively. Some initially wide triplets get disrupted in $t\ga\tau_{\rm coll}/2 \sim 10$ Gyr. } 
\label{fig-2}
\end{figure}

On other hand, however, there are some indications that triplets lie in the periphery of larger systems of galaxies. In a large-scale structure picture, triplets were probably not isolated from tidal perturbations before arriving at their present state. Hence it is of interest to estimate probable tidal effects on the dynamics of triplets in a Hubble time, even though to a first order. In Fig.~\ref{fig-2} we present only the results of the $\sigma$-distribution under a tidal perturbation from a far away `poor cluster'; the triplet was retained at the same initial distance to study the effects of a tidal force.
	The agreement of the median and average $\sigma$ with observations is better when triplets are not considered `island universes'; e.g., $\langle \sigma \rangle \approx 100$ at $t \approx 10$ Gyr when the perturber in Fig.~\ref{fig-2} is at 3 Mpc. 
	Although this value depends obviously on the perturber mass and distance, the results manifest the importance external fields can play in the dynamical properties of otherwise assumed isolated triplets. They also suggests that the environment could have introduced a `selection effect' in allowing some initially wide triplets ($R_{\rm H}\approx 630$ kpc) to become compact, and disrupting others, in a Hubble time.

\section{Final Comments}

Simulations of {\it isolated} triplets show e.g.~that in average rather low velocity dispersions are obtained when compared to observations. In this scenario an underestimate of mass will occur when using the bulk velocity and centre of galaxies, probably by a factor of $\approx 3$. This underestimate can be larger for particular triplets if strong signs of interactions are present and if galaxies have large dark halos;  this situation can be similar for compact groups. Wide triplets could have evolved into compact triplets in a Hubble time, although their high $\sigma$ remains to be explained.

	Tidal perturbations on the evolution of a triplet, however, might have important effects on their dynamics over a Hubble time, and consequently on their mass estimation. We suggest that triplet dynamics is closely tied to cosmology, and that it is not direct to untangle internal effects from external ones. Observational studies that would search e.g.~for possible correlations between $\sigma$ and the density of the triplets environment could shed light on these issues.

\acknowledgments
The author is grateful to the Spanish Ministry of Foreign Affairs for financial support through its MUTIS Program.



\begin{references}
\reference Barnes, J. 1985, \mnras, 215, 517
\reference Karachentsev, I.D. 1999, {\sl these proceedings}
\reference Kuijken, K., \& Dubinski, J. 1995, \mnras, 277, 1341
\reference Lake, G., \& Carlberg, R.G. 1988, AJ, 96, 1587
\reference Nolthenius, R., \& White, S.D.M. 1987, \mnras, 235, 505
\reference Valtonen, M.J., \& Mikkola, S. 1991, ARAA, 29, 9
\reference Zheng, J.Q., Valtonen, M.J., \& Chernin, A.D. 1993, \aj, 105, 2047
\end{references}
\end{document}